\RequirePackage{fix-cm}
\documentclass[smallextended]{svjour3}       
\smartqed  
\usepackage{graphicx}
\usepackage{mathptmx}      
\usepackage{amsmath}
\journalname{}
\begin{document}

\title{The evolution of cooperation through \\institutional incentives and optional participation
}

\titlerunning{Institutional incentives in optional public good games}        

\author{Tatsuya Sasaki         
}


\institute{T. Sasaki \at
              Facluty of Mathematics, University of Vienna, 
              Nordbergstrasse 15, 1090 Vienna, Austria \\
              Tel.: +43-2236-807\\
              Fax: +43-2236-71313 \\
              \email{tatsuya.sasak@univie.ac.at}           \\
	\and	      
	      T. Sasaki \at
              Evolution and Ecology Program, International Institute for Applied Systems Analysis, 
              Schlossplatz 1, 2361 Laxenburg, Austria    
}
\date{Updated: 26 February 2013}

\maketitle

\begin{abstract}
Rewards and penalties are common practical tools that can be used to promote cooperation in social institutions. The evolution of cooperation under reward and punishment incentives in joint enterprises has been formalized and investigated, mostly by using compulsory public good games. Recently, Sasaki {\it et al.} (2012, Proc Natl Acad Sci USA 109:1165--1169) considered optional participation as well as institutional incentives and described how the interplay between these mechanisms affects the evolution of cooperation in public good games. Here, we present a full classification of these evolutionary dynamics. Specifically, whenever penalties are large enough to cause the bi-stability of both cooperation and defection in cases in which participation in the public good game is compulsory, these penalties will ultimately result in cooperation if participation in the public good game is optional. The global stability of coercion-based cooperation in this optional case contrasts strikingly with the bi-stability that is observed in the compulsory case. We also argue that optional participation is not so effective at improving cooperation under rewards.

\keywords{evolutionary game theory \and public good games \and social dilemmas \and rewards \and punishment \and equilibrium selection}
\subclass{91A06 \and 91A22 \and 91A40}
\end{abstract}

\section{Introduction}
Self-interest often leads to freeloading on the contributions of others in the dynamics associated with common goods and joint enterprises [1,2]. As is well known, incentivization, such as rewarding and punishing, is a popular method for harnessing the selfish action and for motivating individuals to behave cooperatively [3--13]. Experimental and theoretical studies on joint enterprises under various incentive schemes are growing [14--22].

Obviously, whether rewards or penalties, sufficiently large incentives can transform freeloaders into full cooperators, and incentives with small impact do nothing on the outcomes [22]. However, incentivizing is costly, and such heavy incentives often incurs serious costs on those who provide the incentives, whether in a peer-to-peer or institutional manner. Previous game-theoretic studies on the evolution of cooperation with incentives have focused on public good games with compulsory participation, and revealed that the intermediate degrees of punishment lead to a couple of stable equilibria, full defection and full cooperation [4,5,10,13,22,23]. In this bi-stable dynamic, establishing full cooperation requires an initially sufficient fraction of cooperators, or ex ante adjustment to overcome the initial condition [10,23]. This situation is a coordination game [24], which is a model of great interest for analyzing a widespread coordination problem (e.g., in choosing distinct technical standards).

In contrast to a traditional case with compulsory participation, another approach to the evolution of cooperation is an option to opt out of joint enterprises [25--37]. The opting-out option can make the freeloader problem relaxed: individuals can exit a joint venture when stuck in a state in which all freeload off one another (``economic stalemate''), and then pursue a stand-alone project; if a joint venture with mutual cooperation is more profitable than in isolation, the individuals once exited will switch to contributing to the venture. This situation, however, will also find defection attractive. Thus, joint enterprises with optional participation can give rise to a rock-paper-scissors cycle [28--31].

Recently, Sasaki {\it et al.} [22] revealed that considering optional participation as well as institutional incentives can effect fully cooperative outcomes for the intermediate ranges of incentives. They demonstrated that opting-out combined with rewarding is not very effective at establishing full cooperation, but opting-out combined with punishment is very effective at establishing cooperation. Although there are a series of existing papers on the interplay of punishment and opting-out mechanisms [38--44], the main points of these earlier studies comprise solving the puzzling issue of second-order freeloading: the exploitation of the efforts of others to uphold incentives for cooperation [2, 4, 7, 45, 46]. Sasaki {\it et al.} [22] consider incentives controlled exclusively by a centralized authority (like the empire or state) [47--50], and thus, their model is already free from the second-order freeloader problem.

Here we analytically provide a full classification of the replicator dynamics in a public good game with institutional incentives and optional participation. We clarify when and how cooperation can be selected over defection in a bi-stable situation associated with institutional punishment without requiring any ability to communicate among individuals. In particular, assuming that the penalties are large enough to cause bi-stability with both full cooperation and full defection (no matter what the basins of attraction are) in cases of compulsory participation, cooperation will necessarily become selected in the long term, regardless of the initial conditions.

\section{Model}
\subsection{Social dilemmas}
To describe our institutional-incentive model, we start from public good games with group size $n \ge 2$. 
The $n$ players in a group are given the opportunity to participate in a public good game.
We assume that participation pays a fixed entrance fee $\sigma > 0$ to the sanctioning institution, whereas non-participation yields nothing. We denote by $m$ the number of players who are willing to participate ($0 \le m \le n$) and assume that at least two participants are required for the game to occur [28,39--42]. If the game does take place, each of the $m$ participants in the group can decide whether to invest a fixed amount $c > 0$ into a common pool, knowing that each contribution will be multiplied by $r > 1$ and then shared equally among all $m - 1$ {\it other} participants in the group. Thus, participants have no direct gain from their own investments [6,41--43,45]. If all of the participants invest, they obtain a net payoff $(r - 1)c > 0$. The game is a social dilemma, which is independent of the value of $r$, because participants can improve their payoffs by withholding their contribution.

Let us next assume that the total incentive stipulated by a sanctioning institution is proportional to the group size $m$ and hence of the form $m\delta$, where $\delta > 0$ is the (potential) per capita incentive. If rewards are employed to incentivize cooperation, these funds will be shared among the so-called ``cooperators'' who contribute (see [51] for a voluntary reward fund). Hence, each cooperator will obtain a bonus that is denoted by $m\delta / n_{\rm C}$, where $n_{\rm C}$ denotes the number of cooperators in the group of $m$ participants. If penalties are employed to incentivize cooperation, ``defectors'' who do not contribute will analogously have their payoffs reduced by $m\delta / n_{\rm D}$, where  $n_{\rm D}$ denotes the number of defectors in the group of $m$ players ($m =n_{\rm C} +n_{\rm D}$).

We consider an infinitely large and well-mixed population of players, from which n samples are randomly selected to form a group for each game. Our analysis of the underlying evolutionary game is based especially on the replicator dynamics [52] for the three corresponding strategies of the cooperator, defector, and non-participant, with respective frequencies $x$, $y$, and $z$. The combination of all possible values of $(x, y, z)$ with $x, y, z \ge 0$ and $x + y + z = 1$ forms the triangular state space $\rm \Delta$. We denote by C, D, and N the three vertices of $\rm \Delta$ that correspond to the three homogeneous states in which all cooperate ($x = 1$), defect ($y = 1$), or are non-participants ($z = 1$), respectively. For $\rm \Delta$, the replicator dynamics are defined by
\begin{equation}
\dot{x}=x(P_{\rm C}^s - \bar{P}^s), \quad \dot{y}=y(P_{\rm D}^s - \bar{P}^s), \quad \dot{z}=z(P_{\rm N}^s - \bar{P}^s),
\end{equation} 
where $\bar{P}^s$ denotes the average payoff in the entire population; $P_{\rm C}^s$, $P_{\rm D}^s$, and $P_{\rm N}^s$ denote the expected payoff values for cooperators, defectors, and non-participants, respectively; and $s = {\rm o}, {\rm r}, {\rm p}$ is used to specify one of three different incentive schemes, namely, ``without incentives,'' ``with rewards,'' and ``with punishment,'' respectively. Because non-participants have a payoff of 0, $P_{\rm N}^s=0$, and thus, $\bar{P}^s=xP_{\rm C}^s+yP_{\rm D}^s$.

We note that if $(r-1)c > \sigma$, the three edges of the state space $\delta$ form a heteroclinic cycle without incentives: N $\to$ C $\to$ D $\to$ N (Figs. 2a or 3a). Defectors dominate cooperators because of the cost of contribution $c$, and non-participants dominate defectors because of the cost of participation $\sigma$. Finally, cooperators dominate non-participants because of the net benefit from the public good game with $(r-1)c > \sigma$. In the interior of $\rm \Delta$, all of the trajectories originate from and converge to N, which is a non-hyperbolic equilibrium. Hence, cooperation can emerge only in brief bursts, sparked by random perturbations [29,41].

\subsection{Pay-offs}
Here, we calculate the average payoff for the whole population and the expected payoff values for cooperators and defectors. In a group with $m - 1$ co-participants ($m = 2, \ldots , n$), a defector or a cooperator obtains from the public good game an average payoff of $rcx/(1-z)$ [41]. Hence,
\begin{equation}
P_{\rm D}^{\rm o} = \left(rc \frac{x}{1-z}-\sigma \right) (1-z^{n-1}).
\end{equation}
Note that $z^{n-1}$ is the probability of finding no co-players and, thus, of being reduced to non-participation. In addition, cooperators contribute $c$ with a probability $1-z^{n-1}$, and thus, $P_{\rm C}^{\rm o} - P_{\rm D}^{\rm o} = -c(1-z^{n-1})$. Hence, $\bar{P}^{\rm o}=(1-z^{n-1})[(r-1)cx-\sigma(1-z)]$.

We now turn to the cases with institutional incentives. First, we consider penalties. Because cooperators never receive penalties, we have $P_{\rm C}^{\rm p}=P_{\rm C}^{\rm o}$. In a group in which the $m-1$ co-participants include $k$ cooperators (and thus, $m-1-k$ defectors), switching from defecting to cooperating implies avoiding the penalty $m\delta / (m-k)$. Hence,
\begin{eqnarray}
P_{\rm C}^{\rm p} - P_{\rm D}^{\rm p} 
&=& (P_{\rm C}^{\rm o} - P_{\rm D}^{\rm o}) + \sum_{m=2}^{n} {n-1 \choose m-1} (1-z)^{m-1} z^{n-m} \nonumber \\
&& \times \left[ 
\sum_{k=0}^{m-1} {m-1 \choose k} \left( \frac{x}{1-z} \right)^k \left( \frac{y}{1-z} \right)^{m-1-k} \frac{m \delta}{m-k} 
\right]
\nonumber \\
&=& -(c-\delta)(1-z^{n-1}) + \delta \frac{x(1-(1-y)^{n-1}}{y},
\end{eqnarray}
and thus,
\begin{eqnarray}
\bar{P}^{\rm p} &=& \bar{P}^{\rm o} - \delta[y(1-z^{n-1}) + x(1-(1-y)^{n-1})] \nonumber \\                                                                 
&=&(1-z^{n-1})((r-1)cx - \sigma(1-z) - \delta y) - \delta x(1-(1-y)^{n-1}).
\end{eqnarray}

Next, we consider rewards. It is now the defectors who are unaffected, implying $P_{\rm D}^{\rm r}=P_{\rm D}^{\rm o}$. In a group with $m-1$ co-participants, including $k$ cooperators, switching from defecting to cooperating implies obtaining the reward $m \delta / (k + 1)$. Hence,
\begin{eqnarray}
P_{\rm C}^{\rm r} - P_{\rm D}^{\rm r} 
&=& (P_{\rm C}^{\rm o} - P_{\rm D}^{\rm o}) + \sum_{m=2}^{n} {n-1 \choose m-1} (1-z)^{m-1} z^{n-m} \nonumber \\
&& \times \left[ 
\sum_{k=0}^{m-1} {m-1 \choose k} \left( \frac{x}{1-z} \right)^k \left( \frac{y}{1-z} \right)^{m-1-k} \frac{m \delta}{k+1} 
\right]
\nonumber \\
&=& -(c-\delta)(1-z^{n-1}) + \delta \frac{y(1-(1-x)^{n-1}}{x},
\end{eqnarray}
and thus,
\begin{eqnarray}
\bar{P}^{\rm r} &=& \bar{P}^{\rm o} + \delta[x(1-z^{n-1}) + y(1-(1-x)^{n-1})] \nonumber \\                                                                                                                                 
&=&(1-z^{n-1})((r-1)cx - \sigma(1-z) + \delta x) - \delta y(1-(1-x)^{n-1}).
\end{eqnarray}

\section{Results}
\subsection{Coordination and coexistence}
We investigated the interplay of institutional incentives and optional participation. As a first step, we considered replicator dynamics along the three edges of the state space $\rm \Delta$. On the DN-edge ($x = 0$), this dynamic is always D $\to$ N because the payoff for non-participating is better than that for defecting by at least the participation fee $\sigma$, regardless of whether penalties versus rewards are in place. On the NC-edge ($y = 0$), it is obvious that if the public good game is too expensive (i.e., if $\sigma \ge (r-1)c$,  under penalties or $\sigma \ge (r-1)c + \delta$, under rewards), players will opt for non-participation more than cooperation. Indeed, N becomes a global attractor because $\dot{z} > 0$ holds in $\rm{\Delta} \setminus \{ z = 0 \}$. We do not consider further cases but assume that the dynamic of the NC-edge is always N $\to$ C.

On the CD-edge ($z = 0$), the dynamic corresponds to compulsory participation, and Eq. (1) reduces to 
$\dot{x} = x(1-x)(P_{\rm C}^s - P_{\rm D}^s)$. Clearly, both of the ends C ($x = 1$) and D ($x = 0$) are fixed points. Under penalties, the term for the payoff difference is
\begin{equation}
P_{\rm C}^{\rm p} - P_{\rm D}^{\rm p} = -c + \delta \frac{1-x^n}{1-x} = -c + \delta \sum_{i=0}^{n-1} x^i .
\end{equation}
Under rewards, it is
\begin{equation}
P_{\rm C}^{\rm r} - P_{\rm D}^{\rm r} = -c + \delta \frac{1-(1-x)^n}{x} = -c + \delta \sum_{i=0}^{n-1} (1-x)^i .
\end{equation}
Because $\delta > 0$, $P_{\rm C}^{\rm r} - P_{\rm D}^{\rm r}$ strictly decreases, and $P_{\rm C}^{\rm p} - P_{\rm D}^{\rm p}$ strictly increases, with $x$. The condition under which there exists an interior equilibrium R on the CD-edge is
\begin{equation}
\delta_- < \delta < \delta_+ , \quad \mathrm{with} \quad \delta_- = \frac{c}{n} \quad \mathrm{and} \quad \delta_+ = c .
\end{equation}

\begin{figure*}
  \includegraphics[width=1.0\textwidth]{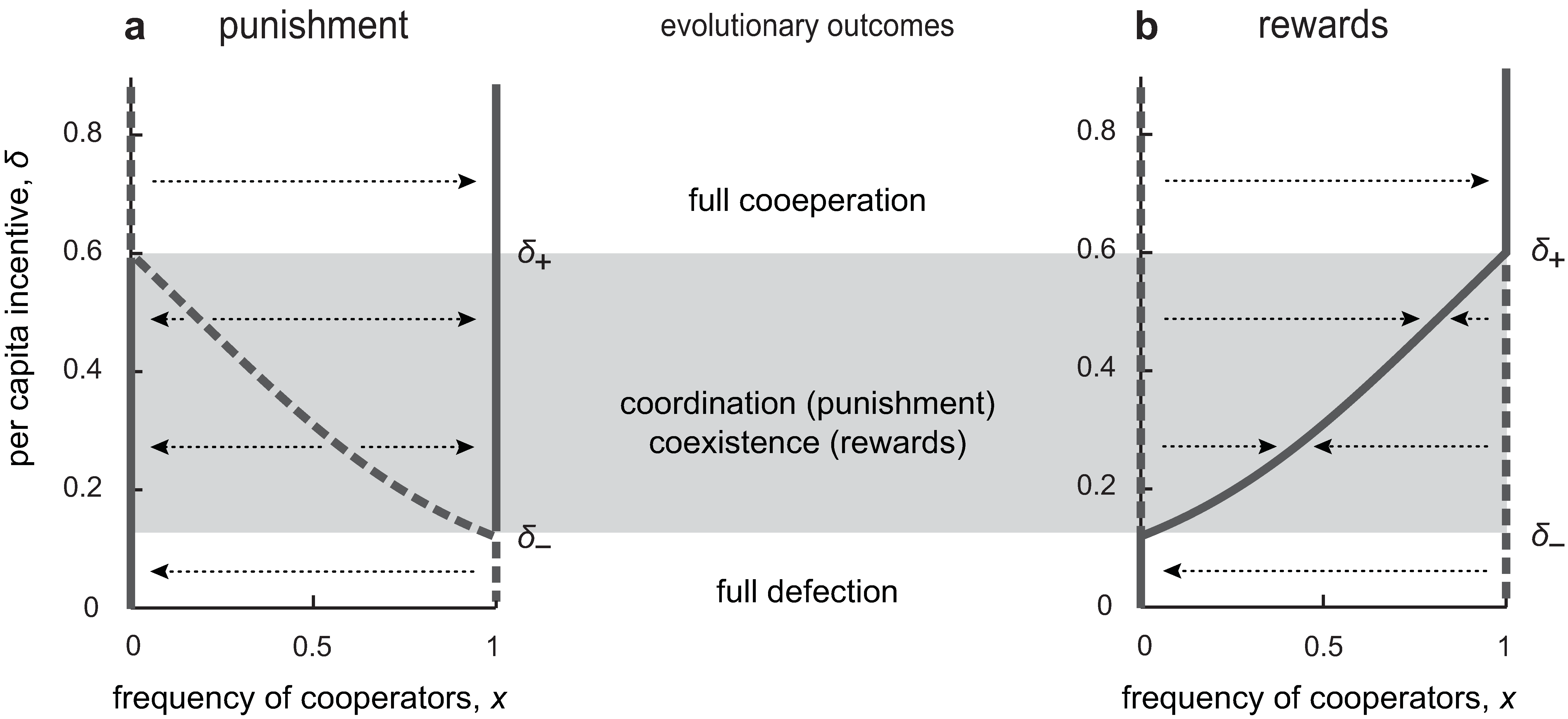}
\caption{
Compulsory public good games with institutional incentives. The location of stable and unstable equilibria (thick continuous lines and dashed lines, respectively) and the direction of evolution (dotted arrows) vary, depending on the per capita incentive, $\delta$. For very small and sufficiently large values of $\delta$, full defection ($x = 0$) and full cooperation ($x = 1$) are the final outcomes, respectively. This applies to both incentives considered. Intermediate values of $\delta$ impact evolutionary dynamics in a strikingly different way, as follows. \textbf{a} Punishment. When $\delta$ increases beyond a threshold $\delta_-$, an unstable interior equilibrium enters the state space at $x = 1$, moves left, and eventually exits it at $x = 0$ for $\delta = \delta_+$. \textbf{b} Rewards. When $\delta$ increases beyond a threshold $\delta_-$, a (globally) stable interior equilibrium enters the state space at $x = 0$, moves right, and eventually exits it at $x = 1$ for $\delta = \delta_+$. Consequently, for the interval $\delta_- < \delta < \delta_+$ (gray-colored region), punishment results in bi-stability of both the pure states; rewards lead to a stable mixture independent of the initial state. Parameters: $n = 5$,  $r = 3$, $c = 1$, and $\sigma = 0.5$.
}
\label{fig:1}       
\end{figure*}

Next, we summarize the game dynamics for compulsory public good games (Fig. 1). For such a small $\delta$ that $\delta < \delta_-$, defection is a unique outcome; D is globally stable, and C is unstable. For such a large $\delta$ that $\delta > \delta_+$, cooperation is a unique outcome; C is globally stable, and D is unstable. For the intermediate values of $\delta$, cooperation evolves in different ways under penalties versus rewards, as follows. Under penalties (Fig. 1a), as $\delta$ crosses the threshold $\delta_-$, C becomes stable, and an unstable interior equilibrium R splits off from C. The point R separates the basins of attraction of C and D. 
Penalties cause bi-stable competition between cooperators and defectors, which is often exhibited as a coordination game [24]; one or the other norm will become established, but there can be no coexistence. 
With increasing $\delta$, the basin of attraction of D becomes increasingly smaller, until $\delta$ attains the value of $\delta_+$. Here, R merges with the formerly stable D, which becomes unstable.
 
In contrast, under rewards (Fig. 1b), as $\delta$ crosses a threshold $\delta_-$, D becomes unstable, and a stable interior equilibrium R splits off from D. The point R is a global attractor. Rewards give rise to the stable coexistence of cooperators and defectors, which is a typical result in a snowdrift game [53]. As $\delta$ increases, the fraction of cooperators within the stable coexistence becomes increasingly larger. Finally, as $\delta$ reaches another threshold $\delta_+$, R merges with the formerly unstable C, which becomes stable. We note that $\delta_+$ and $\delta_-$ have the same value, regardless of whether we take into account rewards or penalties.

\subsection{Uniqueness of the interior equilibrium Q}
Now, we consider the interior of the state space $\rm \Delta$. We start by proving that, for $n > 2$, if an equilibrium Q exists in the interior, it is unique. For this purpose, we introduce the coordinate system $(f, z)$ in $\rm{\Delta} \setminus \{ z = 1 \}$, with $f = x/(x + y)$, and we rewrite Eq. (1) as
\begin{equation}
\dot{f}=f(1-f)(P_{\rm C}^s - P_{\rm D}^s ), \quad \dot{z}=-z \bar{P}^s.
\end{equation}
Dividing the right-hand side of Eq. (10) by $1-z^{n-1}$, which is positive in $\rm{\Delta} \setminus \{ z = 1 \}$, corresponds to a change in velocity and does not affect the orbits in $\rm \Delta$ [52]. Using Eqs. (3)--(6), this transforms Eq. (10) into the following. Under penalties, Eq. (10) becomes
\begin{eqnarray}
\dot{f} &=& f(1-f)[-c + \delta + \delta f H(f,z)], \nonumber \\
\dot{z} &=& z(1-z)[\sigma + \delta - ((r-1)c + \delta)f + \delta f(1-f)H(f,z)],
\end{eqnarray}
whereas under rewards, it becomes
\begin{eqnarray}
\dot{f} &=& f(1-f)[-c + \delta + \delta (1-f) H(1-f,z)], \nonumber \\
\dot{z} &=& z(1-z)[\sigma - ((r-1)c + \delta)f + \delta f(1-f)H(1-f,z)],
\end{eqnarray}
where
\begin{equation}
H(f,z) = \frac{1-[f+(1-f)z]^{n-1}}{(1-f)(1-z^{n-1})} = \frac{1+[f+(1-f)z]+ \cdots +[f+(1-f)z]^{n-2}}{1+z+ \cdots +z^{n-2}}.
\end{equation}
Note that $H(f,0)=\sum_{i=0}^{n-2} f^i$ and $H(f, 1) = 1$.

At an interior equilibrium Q $= (f_{\rm Q}, z_{\rm Q})$, the three different strategies must have equal payoffs, which, in our model, means that they all must equal 0. The conditions $P_{\rm C}^{\rm o}=P_{\rm C}^{\rm p} =0$ under penalties and $P_{\rm D}^{\rm o}=P_{\rm D}^{\rm r}=0$ under rewards imply that $f_{\rm Q}$ is given by
\begin{equation}
f_{\rm Q(p)}=\frac{c+\sigma}{rc} \,\, {\rm under} \,\, {\rm penalties} \,\, {\rm and} \,\, f_{\rm Q(r)}=\frac{\sigma}{rc} \,\, {\rm under} \,\, {\rm rewards}.
\end{equation}
respectively. Thus, if it exists, the interior equilibrium Q must be located on the line given by $f = f_{\rm Q}$. From Eqs. (11) and (12), Q must satisfy 
\begin{equation}
H(f,z)=\frac{c-\delta}{\delta f} \,\, {\rm under} \,\, {\rm penalties} \,\, {\rm and} \,\, H(1-f,z)=\frac{c-\delta}{\delta(1-f)} \,\, {\rm under} \,\, {\rm rewards}.
\end{equation}
In the specific case when $n = 2$, by solving Eqs. (14) and (15) with $H(f,z) = 1$, we can see that the dynamic has an interior equilibrium only when  $\delta = rc^2/((r+1)c+\sigma)$ under penalties or $\delta = rc^2 /(2rc-\sigma)$ under rewards. At this moment, the aforementioned line consists of a continuum of equilibria and connects R and N (Fig. 4). This is a degenerate case of the interior equilibrium, but in Sasaki {\it et al.} [22], this case was not clearly distinguished from the general form described below.

We next show that $z_{\rm Q}$ is uniquely determined in the general case for $n > 2$. Both equations in Eq. (15) have at most one solution with respect to $z$. Because $f_{\rm Q}$ is independent of $z_{\rm Q}$, it is sufficient to show that $H(f, z)$ is strictly monotonic for every $z \in (0,1)$. We first consider penalties. A straightforward computation yields
\begin{eqnarray}
\frac{\partial}{\partial z} H(f,z) &=& \frac{n-1}{(1-f)(1-z^{n-1})^2} [z^{n-2} - (f+(1-f)z)^{n-2} ((1-f)+fz^{n-2})] \nonumber \\
&=& \frac{(n-1) z^{n-2}}{(1-f)(1-z^{n-1})^2}  \nonumber \\
&& \times \left[ 1-\left\{ \left( \frac{f+(1-f)z}{z} \right) ((1-f)+fz) \right\}^{n-2} \frac{(1-f)+fz^{n-2}}{((1-f)+fz)^{n-2}} \right]. \nonumber \\
&&
\end{eqnarray}
We note that
\begin{equation}
 \left( \frac{f+(1-f)z}{z} \right) ((1-f)+fz)
= 1+f(1-f)\left( z-2+\frac{1}{z} \right)
= 1+f(1-f)\frac{(1-z)^2}{z} > 1,
\end{equation}
and
\begin{equation}
 \frac{(1-f)+fz^{n-2}}{((1-f)+fz)^{n-2}} \ge 1.
\end{equation}
This inequality obviously holds for $n = 2$. By induction for every larger $n$, if it holds for $n$, it must hold for $n + 1$ because
\begin{equation}
 \frac{(1-f)+fz^{n+1}}{((1-f)+fz)^{n+1}} - \frac{(1-f)+fz^{n}}{((1-f)+fz)^{n}} 
= \frac{f(1-f)(1-z)(1-z^n)}{((1-f)+fz)^{n+1}} > 0.
\end{equation}
Consequently, the square bracketed term in the last line of Eq. (16) is negative. Thus, $\partial H(f,z) / \partial z < 0$ for every $z \in (0,1)$. We now consider rewards and use the same argument as above. This concludes our proof of the uniqueness of Q.

For $n > 2$, as $\delta$ increases, Q splits off from R (with $x_{\rm R} = f_{\rm Q}$) and moves across the state space along the line given by Eq. (14) and finally exits this space through N. The function $H$ decreases with increasing $z$, and the right-hand side of Eq. (15) decreases with increasing $\delta$, which implies that $z_{\rm Q}$ increases with $\delta$. By substituting Eq. (13) into Eq. (15), we find that the threshold values of $\delta$ for Q's entrance ($z = 0$) and exit ($z = 1$) into the state space are respectively given by
\begin{equation}
\delta_s = \frac{c}{1+B+\cdots+B^{n-1}} \quad {\rm and} \quad \delta^s = \frac{c}{1+B},
\end{equation}
where $B=f_{\rm Q(p)}$ (and $s = \rm{p}$) under penalties, and $B=1-f_{\rm Q(r)}$ (and $s = \rm{r}$) under rewards. We note that $\delta_- < \delta_s \le \delta^s < \delta_+$, which is an equality only for $n = 2$.

\subsection{The saddle point Q}
\label{sec:5}
We next prove that for $n > 2$, Q is a saddle point. We first consider penalties using Eq. (11). Because the square brackets in Eq. (11) vanish at Q, the Jacobian at Q is given by
\begin{equation}
J_{\rm Q} =
\left.
\begin{pmatrix}
\displaystyle \delta f(1-f) \left( H + f  \frac{\partial H}{\partial f} \right)  
\, & \,\, \displaystyle \delta f^2 (1-f) \frac{\partial H}{\partial z} \\ 
& \\
\displaystyle z(1-z) \left[ -A+\delta \left( (1-2f)H+f(1-f) \frac{\partial H}{\partial f} \right) \right] 
\, & \,\, \displaystyle \delta f(1-f)z(1-z) \frac{\partial H}{\partial z}
\end{pmatrix}
\right|_{\rm Q},
\end{equation}
where $H = H(f, z)$ and $A = (r-1)c + \delta$. Using $\partial H(f,z) / \partial z < 0$, $H > 0$, and $A > 0$, which yields
\begin{equation}
\mathrm{det} \, J_{\rm Q} = \delta f^2 (1-f)z(1-z) [A + \delta f H(f,z)] \frac{\partial H(f,z)}{\partial z} < 0.
\end{equation}
Therefore, Q is a saddle point.

We next consider rewards using Eq. (12). Similarly, we find that the Jacobian at Q is given by
\begin{equation}
J_{\rm Q} =
\left.
\begin{pmatrix}
\displaystyle \delta f(1-f) \left( -H + (1-f)  \frac{\partial H}{\partial f} \right)  
\, & \,\, \displaystyle \delta f (1-f)^2 \frac{\partial H}{\partial z} \\ 
& \\
\displaystyle -z(1-z) \left[ A+\delta \left( (1-2f)H+f(1-f) \frac{\partial H}{\partial f} \right) \right] 
\, & \,\, \displaystyle \delta -f(1-f)z(1-z) \frac{\partial H}{\partial z}
\end{pmatrix}
\right|_{\rm Q},
\end{equation}
where $H = H(1-f, z)$ and $A$ is as in Eq. (21). Using $\partial H(1-f,z) / \partial z < 0$, $H > 0$, and $A > 0$, it follows again that $\mathrm{det} \, J_{\rm Q} < 0$. Threrefore, Q is a saddle point.

\subsection{Classification of global dynamics}
\label{sec:6}
Here, we analyze in detail the global dynamics using Eqs. (11) and (12), which are well defined on the entire unit square $U = \{(f, z): 0 \le f \le 1, 0 \le z \le 1 \} $. The induced mapping, $cont:U \to {\rm \Delta}$, contracts the edge $z = 1$ onto the vertex N. Note that ${\rm C} = (1,0)$ and ${\rm D} = (0,0)$ as well as both ends of the edge $z = 1$, ${\rm N}_0 = (0,1)$ and ${\rm N}_1 = (1,1)$, are hyperbolic equilibria, except when each undergoes bifurcation (as shown later). We note that the dynamic on the ${\rm N}_1 {\rm N}_0$-edge is unidirectional to ${\rm N}_0$ without incentives. 

First, we examine penalties. From Eq. (11), the Jacobians at C and ${\rm N}_0$ are respectively given by
\begin{equation}
J_{\rm C} =
　\begin{pmatrix}
　 c-n\delta & 0 \\
　 0 & -[(r-1)c-\sigma]
　\end{pmatrix}
\quad \text{and} \quad
J_{{\rm N}_1} =
　\begin{pmatrix}
　 c-2\delta & 0 \\
　 0 & (r-1)c-\sigma
　\end{pmatrix}.
\end{equation}
From our assumption that $(r-1)c > \sigma$, it follows that if $\delta < c/n$, then $\text{det} J_{\rm C} < 0$, and thus, C is a saddle point; otherwise, $\text{det} J_{\rm C} > 0$ and $\text{tr} J_{\rm C} < 0$, and thus, C is a sink. Regarding ${\rm N}_1$, if $\delta < c/2$, ${\rm N}_1$ is a source ($\text{det} J_{{\rm N}_1} > 0$ and $\text{tr} J_{{\rm N}_1} > 0$); otherwise, ${\rm N}_1$ is a saddle ($\text{det} J_{{\rm N}_1} < 0$). 
Next, the Jacobians at D and ${\rm N}_0$ are respectively given by
\begin{equation}
J_{\rm D} =
　\begin{pmatrix}
　 -\!(c-n\delta) & 0 \\
　 0 & \sigma+\delta
　\end{pmatrix}
\quad \text{and} \quad
J_{{\rm N}_0} =
　\begin{pmatrix}
　 -\!(c-n\delta)  & 0 \\
　 0 & -(\sigma+\delta)
　\end{pmatrix}.
\end{equation}
If $\delta < c$, D is a saddle point ($\text{det}J_{\rm D} < 0$), and ${\rm N}_0$ is a sink ($\text{det}J_{{\rm N}_0} > 0$ and $\text{tr}J_{{\rm N}_0} < 0$); otherwise, D is a source ($\text{det}J_{\rm D} > 0$ and $\text{tr}J_{\rm D} > 0$), and ${\rm N}_0$ is a saddle point ($\text{det}J_{{\rm N}_0} < 0$).

We also analyze the stability of R. As $\delta$ increases from $c/n$ to $c$, the boundary repellor ${\rm R} = (x_{\rm R},0)$ enters the CD-edge at C and then moves to D. The Jacobian at R is given by
\begin{equation}
J_{\rm R} =
\begin{pmatrix}
\displaystyle \delta x_{\rm R} (1-x_{\rm R}) \left. \frac{\partial }{\partial f} f H(f,z) \right|_{\rm R} & \ast \\ 
& \\
0 & -rcx_{\rm R}+(c+\sigma)
\end{pmatrix}.
\end{equation}
Its upper diagonal component is positive because $\partial H(f,z)/ \partial f \ge 0$ and $H > 0$, whereas the lower component vanishes at $x_{\rm R}=f_{\rm Q(p)}=(c + \sigma)/(rc)$. Therefore, if $f_{\rm Q(p)} < x_{\rm R} < 1$, R is a saddle point ($\text{det}J_{\rm R} < 0$) and is stable with respect to $z$; otherwise, if $0 < x_{\rm R} < f_{\rm Q(p)}$, R is a source ($\text{det}J_{\rm R} > 0$ and $\text{tr}J_{\rm D} > 0$).

In addition, a new boundary equilibrium ${\rm S} = (x_{\rm S},1)$ can appear along the ${\rm N}_1 {\rm N}_0$-edge. Solving $\dot{f}(x_{\rm S},1)=0$ in Eq. (11) yields $x_{\rm S} = (c-\delta) / \delta$; thus, S is unique. S is a repellor along the edge (as is R). As $\delta$ increases, S enters the edge at ${\rm N}_1$ (for $\delta = c/2$) and exits it at ${\rm N}_0$ (for $\delta = c$). The Jacobian at S is given by
\begin{equation}
J_{\rm S} =
\begin{pmatrix}
\displaystyle \delta x_{\rm S} (1-x_{\rm S}) \left. \frac{\partial }{\partial f} f H(f,z) \right|_{\rm S} & \ast \\ 
& \\
0 & \delta x_{\rm S}^2+(r-1)cx_{\rm S}-\sigma-\delta
\end{pmatrix}.
\end{equation}
Again, its upper diagonal component is positive. Using $x_{\rm S} = (c-\delta)/ \delta$, we find that the sign of the lower component changes once, from positive to negative, as $\delta$ increases from $c/2$ to $c$. Therefore, S is initially a source ($\text{det}J_{\rm S} > 0$ and $\text{tr}J_{\rm S} > 0$) but then turns into a saddle point ($\text{det}J_{\rm S} < 0$), which is stable with respect to $z$.

\begin{figure*}
  \includegraphics[width=1.0\textwidth]{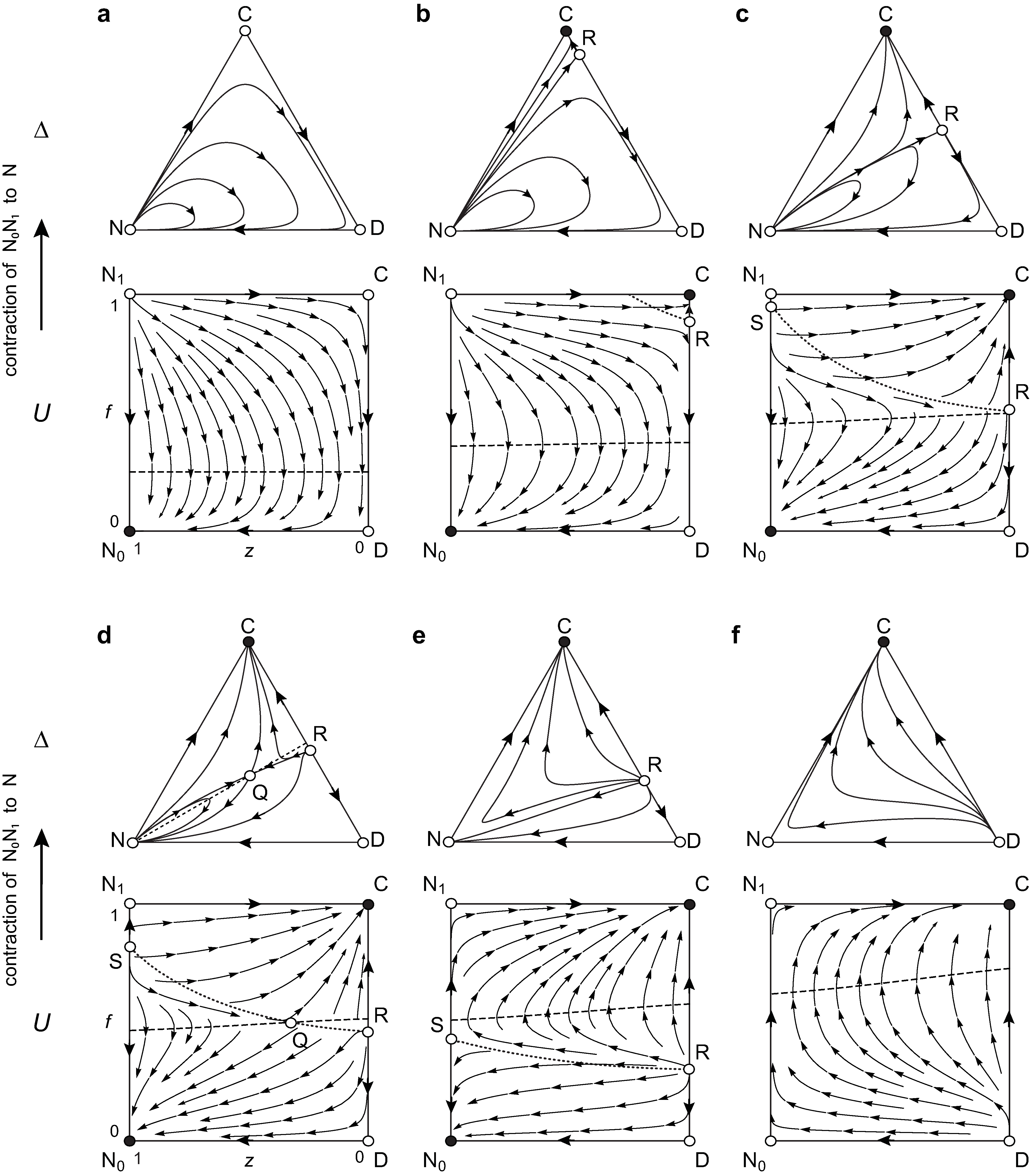}
\caption{
Optional public good games with institutional punishment. The triangles represent the state space ${\rm \Delta} = \{ (x, y, z): x, y, z > 0, x + y + z = 1 \}$. Its vertices C, D, and N correspond to the three homogeneous states of cooperators ($x = 1$), defectors ($y = 1$), and non-participants ($z = 1$), respectively. The unit squares represent an extended state space $U = \{(f, z): 0 \le f \le 1, 0 \le z \le 1 \}$ such that  $\rm \Delta$ is its image according to the mapping $x = f(1-z)$, $y = (1-f)(1-z)$, which is injective except for $z = 1$. The edge is contracted to N. The vertices of $U$ are denoted by ${\rm C} = (1,0)$, ${\rm D} = (0,0)$, ${\rm N}_1 = (1,1)$, and ${\rm N}_0 = (0,1)$. The stream plot is based on Eq. (11). Dot and dashed curves in $U$ denote where $\dot{f}$ and $\dot{z}$ vanish, respectively. \textbf{a} Without incentives, the interior of $U$ is filled with orbits originating from ${\rm N}_1$ and then converging to ${\rm N}_0$, which correspond to homoclinic cycles to fully cover the interior of  $\rm \Delta$. \textbf{b} As $\delta$ increases, the equilibrium R (a saddle point) first enters the CD-edge at C, which then becomes a sink. \textbf{c} When $\delta$ crosses $c/2$, the equilibrium S (a source) enters the ${\rm N}_1 {\rm N}_0$-edge at ${\rm N}_1$, which then becomes a saddle point. \textbf{d} When $\delta$ crosses $\delta_{\rm p}$, the saddle point Q enters the interior of  $\rm \Delta$ through R, which then becomes a source. Q traverses $U$ along a horizontal line. \textbf{e} When $\delta$ crosses $\delta_{\rm p}$, Q exits  $\rm \Delta$ through S, which then becomes a saddle. For larger values of $\delta$, there is no such interior orbit that originates from the ${\rm N}_1 {\rm N}_0$-edge and converges to it, and thus, $\rm \Delta$ has no homoclinic cycle. When $\delta$ crosses $\delta_+$, R and S exit  $\rm \Delta$ through D, which becomes a source, and ${\rm N}_0$, which becomes a saddle. \textbf{f} For $\delta > \delta_+$, the interiors of $U$ and  $\rm \Delta$ are filled with orbits originating from D and converging to C. Parameters are the same as in Fig. 1: $n = 5$,  $r = 3$, $c = 1$, $\sigma = 0.5$, and $\delta = 0$ (a), 0.25 (b), 0.51 (c), 0.55 (d), 0.7 (e), or 1.2 (f) }
\label{fig:2}       
\end{figure*}

We give a full classification of the global dynamics under penalties, as follows.
\begin{enumerate}
\item For $0 \le \delta < \delta_-$ (Fig. 2a), C and D are saddle points, ${\rm N}_1$ is a source, and ${\rm N}_0$ is a sink. There is no other equilibrium, and $\dot{f} < 0$ holds in the interior state space. All interior orbits originate from ${\rm N}_1$ and converge to ${\rm N}_0$. ${\rm N}_0$ is globally stable. After applying the contraction map, we find that the interior of $\rm \Delta$ is filled with homoclinic orbits originating from and converging to N.
 
\item As $\delta$ crosses $\delta_-$ (Fig. 2b), C becomes a sink, and the equilibrium R enters the CD-edge at C. R is unstable along that edge but is stable with respect to $z$. Therefore, there is an orbit originating from ${\rm N}_1$ and converging to R that separates the basins of attraction of C and ${\rm N}_0$. All of the orbits in the basin of ${\rm N}_0$ have their $\alpha$-limits at ${\rm N}_1$, as before. Hence, the corresponding region in $\rm \Delta$ is filled with homoclinic orbits and is surrounded by a heteroclinic cycle N $\to$ R $\to$ D $\to$ N. However, if the population is in the vicinity of N, small and rare random perturbations will eventually send the population into the basin of attraction of C (as is the case for $c/2 < \delta$).

\item As $\delta$ crosses $c/2$ (Fig. 2c), ${\rm N}_1$ becomes a saddle point, and a new equilibrium S enters the ${\rm N}_1 {\rm N}_0$-edge at ${\rm N}_1$. S is a source. As $\delta$ increases, S moves toward ${\rm N}_0$. If $c/2 < \delta_{\rm p}$ holds, then for $c/2 < \delta < \delta_{\rm p}$, there is still an orbit originating from S and converging to R that separates the state space into basins of attraction of C and ${\rm N}_0$. All of the orbits in the basin of ${\rm N}_0$ have their $\alpha$-limits at ${\rm N}_1$, as before. In $\rm \Delta$, the separatrix NR and the NC-edge now intersect transversally at N, and the entrance of a minority of participants (including cooperators and defectors) into the greater population of non-participants may be successful.

\item As $\delta$ crosses $\delta_{\rm p}$ (Fig. 2d), the saddle point Q enters the interior of $\rm \Delta$ through R, which becomes a source. Based on the uniqueness of Q and the Poincar{\rm \'{e}}-Bendixson theorem ([52], Appendix A), we can see that there is no such homoclinic orbit originating from and converging to Q, and the unstable manifold of Q must consist of an orbit converging to C and an orbit converging to ${\rm N}_0$; the stable manifold of Q must consist of an orbit originating from D and an orbit originating from S (or, in the case that $\delta_{\rm p} < c/2$, from ${\rm N}_1$ for $\delta_{\rm p} < \delta < c/2$). The stable manifold separates the basins of attraction of C and ${\rm N}_0$; the unstable manifold separates the basin for ${\rm N}_0$ into two regions. One of them is filled with orbits originating from S (or from ${\rm N}_1$ under the above conditions) and converging to ${\rm N}_0$. For $\rm \Delta$, this means that the corresponding region is filled with homoclinic orbits and is surrounded by a heteroclinic cycle N $\to$ Q $\to$ N (Fig. 2d). As $\delta$ further increases, Q moves across $U$, from the CD-edge to the ${\rm N}_1 {\rm N}_0$-edge along the line $f = f_{\rm Q(p)}$. For $n = 2$, R and S undergo bifurcation simultaneously, and the linear continuum of interior equilibria, which connects R and S, appears only at the bifurcation point (Fig. 4a). 

\item As $\delta$ crosses $\delta^{\rm p}$ (Fig. 2e), Q exits the state space through S, which then becomes saturated. For larger values of $\delta$, there is no longer an interior equilibrium. S is a saddle point, which is connected with the source R by an orbit leading from R to S.    

\item Finally, as $\delta$ crosses $\delta_+$ (Fig. 2f), R and S simultaneously exit $U$, through D and ${\rm N}_0$, respectively. For $\delta_+ < \delta$, ${\rm N}_1$ and ${\rm N}_0$ are saddle points, D is a source, and C is a sink. $\dot{f} > 0$ holds throughout the state space. All of the interior orbits originate from D and converge to C. Hence, C is globally stable.   
\end{enumerate}

Let us now turn to rewards. From Eq. (12), the Jacobians at D and ${\rm N}_0$ are

\begin{equation}
J_{\rm D} =
　\begin{pmatrix}
　 -\!(c-n\delta) & 0 \\
　 0 & \sigma
　\end{pmatrix}
\quad \text{and} \quad
J_{{\rm N}_0} =
　\begin{pmatrix}
　 -\!(c-2\delta) & 0 \\
　 0 & -\sigma
　\end{pmatrix}.
\end{equation}
If $\delta < c/n$, D is a saddle point ($\text{det}J_{\rm D} < 0$); otherwise, D is a source ($\text{det}J_{\rm D} > 0$ and $\text{tr}J_{\rm D} > 0$). Regarding ${\rm N}_0$, if $\delta < c/2$, ${\rm N}_0$ is a sink ($\text{det}J_{{\rm N}_0} > 0$ and $\text{tr}J_{{\rm N}_0} < 0$); otherwise, ${\rm N}_0$ is a saddle point ($\text{det}J_{{\rm N}_0} < 0$). Meanwhile, the Jacobians at C and ${\rm N}_1$ are
\begin{equation}
J_{\rm C} =
　\begin{pmatrix}
　 c-\delta & 0 \\
　 0 & -[(r-1)c-\sigma+\delta]
　\end{pmatrix}
\quad \text{and} \quad
J_{{\rm N}_1} =
　\begin{pmatrix}
　 c-\delta & 0 \\
　 0 & (r-1)c-\sigma+\delta
　\end{pmatrix}.
\end{equation}
From $(r-1)c > \sigma - \delta$, it follows that if $\delta < c$, C is a saddle point ($\text{det}J_{\rm C} < 0$), and ${\rm N}_1$ is a source ($\text{det}J_{{\rm N}_1} > 0$ and $\text{tr}J_{{\rm N}_1} > 0$); otherwise, C is a sink ($\text{det}J_{\rm C} > 0$ and $\text{tr}J_{\rm C} < 0$), and ${\rm N}_1$ is a saddle point ($\text{det}J_{{\rm N}_1} < 0$).

We also analyze the stability of R. As $\delta$ increases from $c/n$ to $c$, the boundary attractor R enters the CD-edge at D and then moves toward C. The Jacobian at R is given by
\begin{equation}
J_{\rm R} =
\begin{pmatrix}
\displaystyle -\delta x_{\rm R} (1-x_{\rm R}) \left. \frac{\partial }{\partial f} (1-f) H(1-f,z) \right|_{\rm R} & \ast \\ 
& \\
0 & -rcx_{\rm R}+\sigma
\end{pmatrix}.
\end{equation}
Its upper diagonal component is negative because $\partial H(1-f,z)/ \partial f \le 0$ and $H > 0$, and the lower component vanishes at $x_{\rm R} = f_{\rm Q(r)} = \sigma / (rc)$. Therefore, if $0 < x_{\rm R} < f_{\rm Q(r)}$, R is a saddle point ($\text{det}J_{\rm R} < 0$) and unstable with respect to $z$; otherwise, if $f_{\rm Q(r)} < x_{\rm R} < 1$, R is a sink ($\text{det}J_{\rm R} > 0$ and $\text{tr}J_{\rm R} < 0$).

Similarly, a boundary equilibrium S can appear along the ${\rm N}_1 {\rm N}_0$-edge. Solving $\dot{f}(x_{\rm S},1)=0$ in Eq. (12) yields $x_{\rm S} = 1 - (c - \delta) / \delta$, and thus, S is unique. S is an attractor along the edge (as is R). As $\delta$ increases, S enters the edge at ${\rm N}_0$ (for $\delta = c/2$) and exits at ${\rm N}_1$ (for $\delta = c$). The Jacobian at S is
\begin{equation}
J_{\rm S} =
\begin{pmatrix}
\displaystyle -\delta x_{\rm S} (1-x_{\rm S}) \left. \frac{\partial }{\partial f} (1-f) H(1-f,z) \right|_{\rm S} & \ast \\ 
& \\
0 & -[\delta x_{\rm S}^2 - ((r-1)c+2\delta)x_{\rm S} + \sigma]
\end{pmatrix}.
\end{equation}
Again, its upper diagonal component is positive. Using $x_{\rm S} = 1 - (c - \delta) / \delta$, we find that the sign of the lower component changes once, from negative to positive, as $\delta$ increases from $c/2$ to $c$. Therefore, S is initially a sink ($\text{det}J_{\rm S} > 0$ and $\text{tr}J_{\rm S} < 0$) and then becomes a saddle point ($\text{det}J_{\rm S} < 0$), which is unstable with respect to $z$.

\begin{figure*}
  \includegraphics[width=1.0\textwidth]{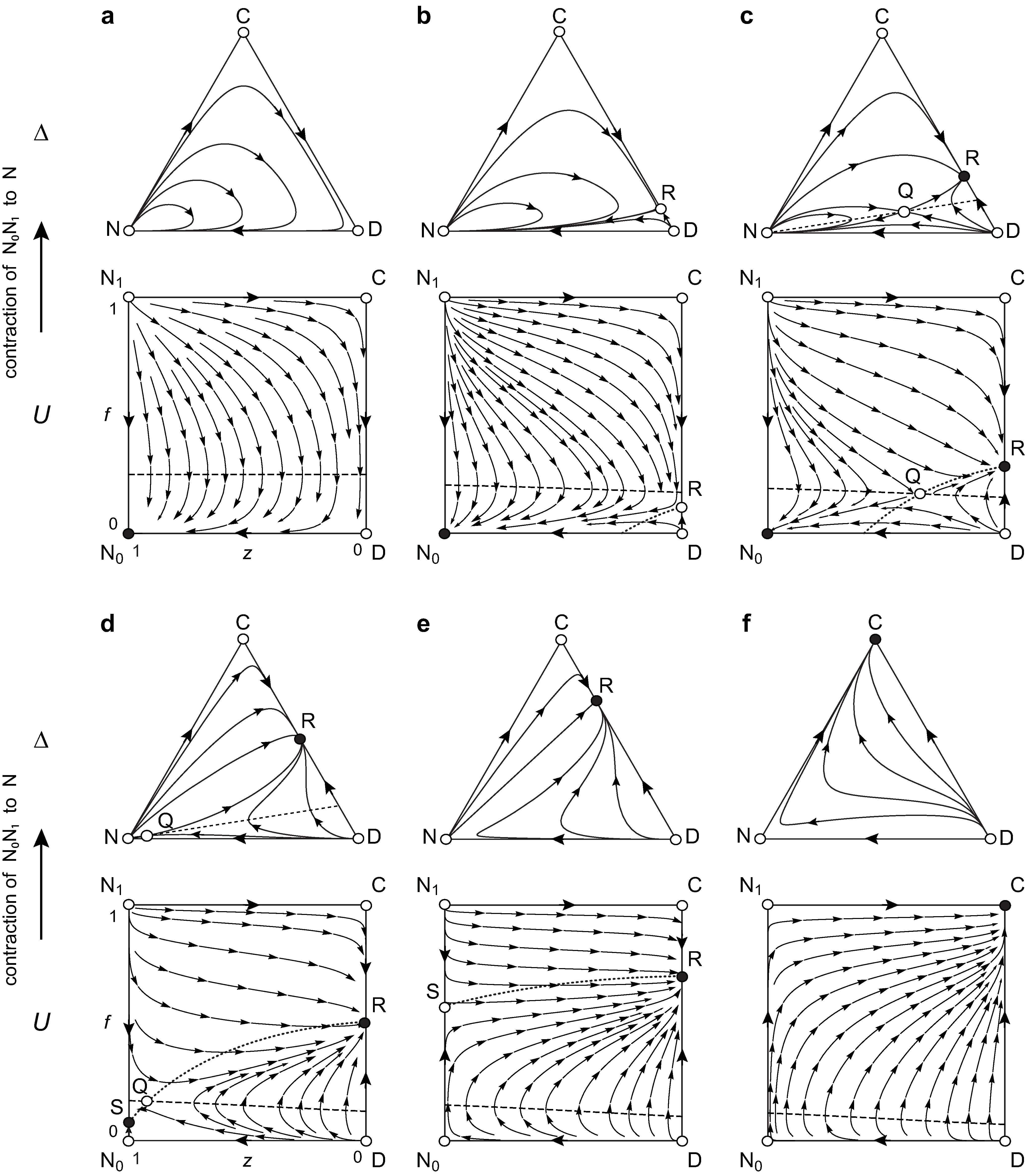}
\caption{
Optional public good games with institutional rewards. Notations are as in Fig. 2, and the stream plot is based on Eq. (12). \textbf{a} Without incentives, this figure is same as Fig. 1a. \textbf{b} As $\delta$ increases, the equilibrium R (a saddle point) first enters the CD-edge at D, which then becomes a source. \textbf{c} When $\delta$ crosses $\delta_{\rm r}$, the saddle point Q enters the interior of $\rm \Delta$ through R, which then becomes a sink. Q traverses $U$ along a horizontal line. \textbf{d} When $\delta$ crosses $c/2$, the rest point S (a sink) enters the ${\rm N}_1 {\rm N}_0$-edge at ${\rm N}_0$, which then becomes a saddle point. \textbf{e} When $\delta$ crosses $\delta^{\rm r}$, Q exits $U$ through S, which then becomes a saddle point. For larger values of $\delta$, there is no such interior orbit that originates from the ${\rm N}_1 {\rm N}_0$-edge and converges to it and, thus, $\rm \Delta$ has no homoclinic cycle. When $\delta$ crosses $\delta_+$, R and S exit $\rm \Delta$ through C, which becomes a sink, and ${\rm N}_1$, which becomes a saddle. \textbf{f} For $\delta > \delta_+$, the interiors of $U$ and $\rm \Delta$ are filled with orbits originating from D and then converging to C, as in the case with institutional punishment. The parameters are the same as in Figs. 1 and 2, except $\delta=0$ (a), 0.25 (b), 0.35 (c), 0.52 (d), 0.7 (e), or 1.2 (f) }
\label{fig:3}       
\end{figure*}

A full classification of the global dynamics under rewards is as follows.
\begin{enumerate}
\item For $0 \le \delta < \delta_-$ (Fig. 3a), C and D are again saddle points, ${\rm N}_1$ is a source, and ${\rm N}_0$ is a sink. $\dot{f} < 0$ holds in the interior state space, and all of the interior orbits originate from ${\rm N}_1$ and converge to ${\rm N}_0$. The interior of $\rm \Delta$ is filled with homoclinic orbits originating from and converging to N.

\item As $\delta$ crosses $\delta_-$ (Fig. 3b), D turns into a source, and the saddle point R enters the CD-edge through D. There exists an orbit originating from R and converging to ${\rm N}_0$. In contrast to the case with penalties, ${\rm N}_0$ remains a global attractor. A region separated by the orbit R${\rm N}_0$ encloses orbits with ${\rm N}_1$ as their $\alpha$-limit. Therefore, in $\rm \Delta$, the corresponding region is filled with homoclinic orbits that are surrounded by a heteroclinic cycle N $\to$ C $\to$ R $\to$ N.

\item As $\delta$ crosses $c/2$, ${\rm N}_0$ becomes a saddle point, and the equilibrium S enters the ${\rm N}_1 {\rm N}_0$-edge at ${\rm N}_0$. S is a sink (and thus, a global attractor). As $\delta$ increases, S moves to ${\rm N}_1$. If $c/2 < \delta_{\rm r}$ holds, then for $c/2 < \delta < \delta_{\rm r}$, there exists an orbit originating from R and converging to S, which separates the interior state space into two regions. One of these regions consists of orbits originating from ${\rm N}_1$, corresponding in $\rm \Delta$ to a region filled with homoclinic orbits. The other region consists of orbits originating from D. In $\rm \Delta$, the separatrix RN and the NC-edge intersect transversally at N.

\item As $\delta$ crosses $\delta_{\rm r}$ (Fig. 3d), the saddle point Q enters the interior state space through R, which then becomes a sink. There is no homoclinic loop for Q, as before, and now, we find that the stable manifold of Q must consist of two orbits originating from D and ${\rm N}_1$. The unstable manifold of Q must consist of an orbit converging to R and another converging to S or, in the case that $\delta_{\rm r} < c/2$, converging to ${\rm N}_0$ for $\delta_{\rm r} < \delta < c/2$ (Fig. 3c). The stable manifold separates the basins of attraction of R and S (or ${\rm N}_0$ under the above conditions); the unstable manifold separates the basin for S (or ${\rm N}_0$) into two regions. One of these regions is filled with orbits issuing from ${\rm N}_1$ and converging to S (or ${\rm N}_0$). The corresponding region in $\rm \Delta$ is filled with homoclinic orbits and is surrounded by a heteroclinic cycle N $\to$ Q $\to$ N (Figs. 3c and 3d). As $\delta$ continues to increase, Q moves through $U$, from the CD-edge to ${\rm N}_1 {\rm N}_0$, along the line $f = f_{\rm Q(r)}$. For $n = 2$, R and S undergo bifurcation simultaneously, and the continuum of interior equilibria, which connects R and S, appears only at the bifurcation point (Fig. 4b).

\item As $\delta$ crosses $\delta^{\rm r}$ (Fig. 3e), Q exits the state space through S, which then becomes a saddle point. For larger values of $\delta$, there is no longer an interior equilibrium. S is connected with the sink R by an orbit from S to R. All of the interior orbits converge to R.

\item Finally, as $\delta$ crosses $\delta_+$ (Fig. 3f), R and S simultaneously exit $U$ through C and ${\rm N}_1$, respectively. Just as in the case with punishment, for $\delta_+ < \delta$, ${\rm N}_1$ and ${\rm N}_0$ are saddle points, and D is a source. Finally, C is a sink. $\dot{f} > 0$ holds throughout the state space. All of the interior orbits originate from D and then converge to C. Hence, C is globally stable. 
\end{enumerate}

\begin{figure*}
  \includegraphics[width=0.7\textwidth]{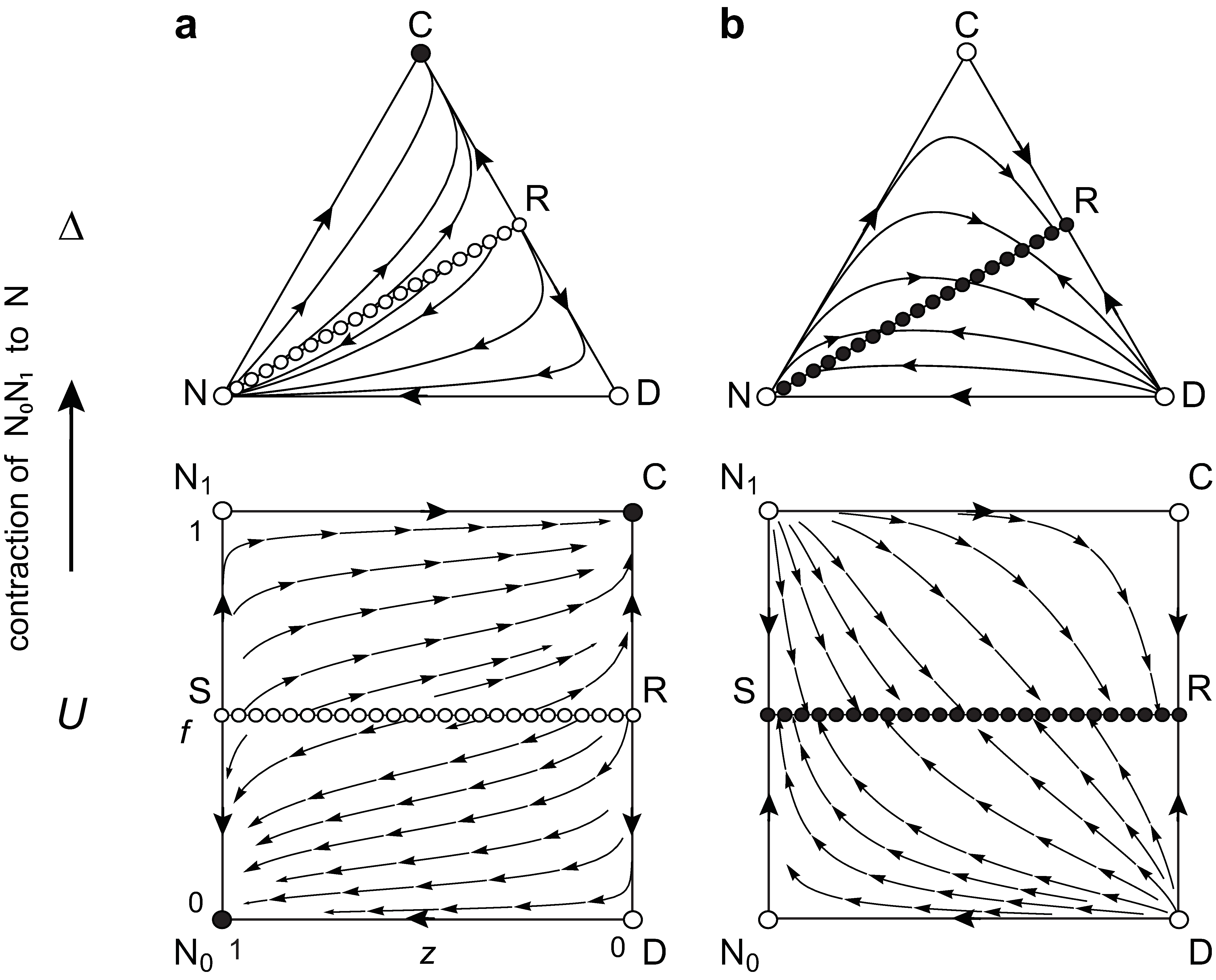}
\caption{
Optional public good games with institutional incentives for $n = 2$. Notations are as in Fig. 2, and the stream plot is given under (a) penalties based on Eq. (11) and (b) rewards based on Eq. (12). Other parameters include $c = 1$, $\sigma = 0.5$, and $r = 3$ (a) or 1 (b). For $n = 2$, the state space may have an interior equilibrium, which is a linear continuum of the equilibria, only at $\delta = 2/3$. \textbf{a} Under penalties, the fixed line that connects N (S in $U$) and R is repelling with respect to f and divides $\rm \Delta$ into basins of attraction of N (${\rm N}_0$ in $U$) and C. From the vicinity of N, arbitrarily small random perturbations will send the state into the region of attraction of C. \textbf{b} Under rewards, the fixed line is attracting with respect to $f$, and thus, the interior orbits converge to corresponding points on the line}
\label{fig:4}       
\end{figure*}

\section{Discussion}
We considered a model for the evolution of cooperation through institutional incentives and analyzed in detail evolutionary game dynamics. Specifically, based on a public good game with optional participation, we fully analyzed how opting-opt impacts game dynamics; in particular, opting-out can completely relax a coordination problem associated with punishment for a considerably broader range of parameters than in cases of compulsory participation.

We start from assuming that there is a state-like institution that takes exclusive control of individual-level sanctions in the form of penalties and rewards. In our extended model, nobody is forced to enter a joint enterprise that is protected by the institutional sanctioning, however, whoever is willing to enter, must be charged at the entrance. Further, if one proves unable or unwilling to pay, the sanctioning institution can ban that person from participation in the game. Indeed, joint ventures in real life are mostly protected by enforceable contracts in which members can freely participate, but are bound by a higher authority. For example, anyone can opt to not participate in a wedding vow (with donating to the temple or church), but once it is taken, it is the strongest contract among enforceable contracts. As far as we know, such higher authorities always demand penalties if contracts are broken.

Based on our mathematical analysis, we argue that institutional punishment, rather than institutional rewards, can become a more viable incentivization scheme for cooperation when combined with optional participation. We show that combining optional participation with rewards can complicate the game dynamics, especially if there is an attractor with all three strategies: cooperation, defection, and non-participation, present. This can only marginally improve group welfare for a small range of per capita incentive $\delta$, with $\delta_- < \delta < \delta_{\rm r}$ (Fig. 3b). Within this interval, compulsory participation can lead to partial cooperation; however, optional participation eliminates the cooperation and thus drives a population into a state in which all players exit. Hence, freedom of participation is not a particularly effective way of boosting cooperation under a rewards scenario.

Under penalties, the situation varies considerably. Indeed, as soon as $\delta > \delta_-$ (Fig. 2b), the state in which all players cooperate abruptly turns into a global attractor for optional participation. When $\delta$ just exceeds $\delta_-$, group welfare becomes maximum $(r - 1)c - \sigma$.  Meanwhile, for compulsory participation, almost all of the (boundary) state space between cooperation and defection still belongs to the basin of attraction of the state in which all players defect. Because $\delta = c/n$, where $n$ is the group size, and $c$ is the net contribution cost (a constant), when $n$ is larger, the minimal institutional sanctioning cost $\delta_-$ to establish full cooperation is smaller.

There are various approaches to equilibrium selection in $n$-person coordination games for binary choices [54--56]. A strand of literature bases stochastic evolution models [57--59], in which typically, a “risk-dominant” equilibrium [60] that has the larger basin of attraction is selected through random fluctuation in the long run. In contrast, considering optional participation, our model typically selects the cooperative equilibrium which provides the higher group welfare, even if the cooperative equilibrium has the smaller basin of attraction when participation is compulsory than has the defective equilibrium. In the sense of favoring the efficient equilibrium, our result is similar to that found in a decentralized partner-changing model proposed by Oechssler [61], in which players may occasionally change interaction groups.

Throughout centralized institutional sanctions mentioned so far, norm-based cooperation is less likely to suffer from higher-order freeloaders, which have been problematic in modeling decentralized peer-to-peer sanctions [2,62]. In addition, it is clear that sanctioning institutions will stipulate a lesser antisocial punishment targeted at cooperators [63], which can prevent the evolution of pro-social behaviors ([64,65], see also [36]). Indeed, punishing cooperators essentially promote defectors, who will reduce the number of participants willing to pay for social institutions. For self-sustainability, thus, sanctioning institutions should dismiss any antisocial schemes that may lead to a future reduction in resources for funding the institution.
 
Thus, we find that our model reduces the space of possible actions into a very narrow framework of alternative strategies, in exchange for increasing the degree of the institution’s complexity and abstractiveness. In practice, truly chaotic situations which offer a very long list of possibilities are unfeasible and create inconvenience, as is described by Michael Ende in ``{\it The Prison of Freedom}'' [1992]. Participants in all economic experiments usually can make their meaningful choices only in a short and regulated list of options, as is the way with us in real life. Our result indicates that a third party capable of exclusively controlling incentives and membership can play a key role in selecting a cooperative equilibrium without ex ante adjustment. The question of how such a social order can emerge out of a world of chaos is left entirely open.

\begin{acknowledgements}
We thank {\AA}ke Br\"{a}nnstr\"{o}m, Ulf Dieckmann, and Karl Sigmund for their comments and suggestions on an earlier version of this paper. This study was enabled by financial support by the FWF (Austrian Science Fund) to Ulf Dieckmann at IIASA (TECT I-106 G11), and was also supported by grant RFP-12-21 from the Foundational Questions in Evolutionary Biology Fund. 
\end{acknowledgements}

\section*{Appendix A}
First, we prove that a homoclinic loop that originates from and converges to Q does not exist. Using the Poincar{\rm \'{e}}-Bendixson theorem [52] and the uniqueness of an interior equilibrium, we show that if it does exist, there must be a point $p$ inside the loop such that both of its $\alpha$- and $\omega$-limit sets include Q. This contradicts the fact that Q is a saddle point. Indeed, there may be a section that cuts through Q such that the positive and negative orbits of $p$ infinitely often cross it; however, it is impossible for a sequence consisting of all the crossing points to originate from and also converge to the saddle point Q. Hence, there is no homoclinic orbit of Q.

Next, we show that orbits that form the unstable manifold of Q do not converge to the same equilibrium (indeed, this is a sink). If they do, the closed region that is surrounded by the orbits must include a point $q$ such that its $\omega$-limit set is Q. Using the Poincar{\rm \'{e}}-Bendixson theorem and the uniqueness of an interior equilibrium, the $\alpha$-limit set for $q$ must include Q; this is a contradiction. Similarly, we can prove that the orbits that form the stable manifold of Q do not issue from the same equilibrium.


\begin{thebibliography}{}
%
%
\bibitem{Har68}
Hardin G (1968) The tragedy of the commons. Science 162:1243--1248. 
doi:10.1126/science.162.3859.1243
\bibitem{Ost90}
Ostrom E (1990) Governing the Commons: The Evolution of Institutions for Collective Action. Cambridge University Press, New York
\bibitem{Ols65}
Olson E (1965) The Logic of Collective Action: Public Goods and the Theory of Groups Harvard University Press, Cambridge, MA
\bibitem{BoyRic92}
Boyd R, Richerson P (1992) Punishment allows the evolution of cooperation (or anything else) in sizable groups. Ethol Sociobiol 13:171--195.
doi:10.1016/0162-3095(92)90032-Y
\bibitem{SigEtal01}
Sigmund K, Hauert C, Nowak MA (2001) Reward and punishment. Proc Natl Acad Sci USA 98:10757--10762.
doi: 10.1073/pnas.161155698
\bibitem{FehGae02}
Fehr E, G\"{a}chter S (2000) Cooperation and punishment in public goods experiments. Am Econ Rev 90:980--994. 
doi:10.1257/aer.90.4.980
\bibitem{Oli80}
Oliver P (1980) Rewards and punishments as selective incentives for collective action: theoretical investigations. Am J Sociol 85:1356--1375.
doi:10.1086/227168
\bibitem{Sig07}
Sigmund K (2007) Punish or perish? Retaliation and collaboration among humans. Trends Ecol Evol 22:593--600.
doi:10.1016/j.tree.2007.06.012
\bibitem{RanEtal09}
Rand DG, Dreber A, Ellingsen T, Fudenberg D, Nowak MA (2009) Positive interactions promote public cooperation. Science 325:1272--1275.
doi:10.1126/science.1177418
\bibitem{BoyEtal10}
Boyd R, Gintis H, Bowles S (2010) Coordinated punishment of defectors sustains cooperation and can proliferate when rare. Science 328:617--620.
doi:10.1126/science.1183665
\bibitem{BMV11}
Balliet D, Mulder LB, Van Lange PAM (2011) Reward, punishment, and cooperation: a meta-analysis. Psychol Bull 137:594--615. 
doi:10.1037/a0
\bibitem{Gaech12}
G\"{a}chter S (2012) Social science: carrot or stick? Nature 483:39--40.
doi:10.1038/483039a
\bibitem{SasUch13}
Sasaki T, Uchida S (2013) The evolution of cooperation by social exclusion. Proc R Soc B 280:1752. 
doi:10.1098/rspb.2012.2498
\bibitem{AndEtal03}
Andreoni J, Harbaugh WT, Vesterlund L (2003) The carrot or the stick: rewards, punishments, and cooperation. Am Econ Rev 93:893--902.
doi:10.1257/000282803322157142
\bibitem{GueEtal06}
G\"{u}rerk O, Irlenbush B, Rockenbach B (2006) The competitive advantage of sanctioning institutions. Science 312:108--111.
doi:10.1126/science.1123633
\bibitem{Sefton07}
Sefton M, Shupp R, Walker JM. (2007) The effect of rewards and sanctions in provision of public goods. Econ inq 45:671--690. 
doi:10.1111/j.1465-7295.2007.00051.x
\bibitem{GueEtal09}
G\"{u}rerk O, Irlenbusch B, Rockenbach B (2009) Motivating teammates: the leader’s choice between positive and negative incentives. J Econ Psychol 30:591--607.
doi:10.1016/j.joep.2009.04.004
\bibitem{OgoEtal09}
O'Gorman R, Henrich J, Van Vugt M. (2009) Constraining free riding in public goods games: designated solitary punishers can sustain human cooperation. Proc R Soc B 276:323--329.
doi:10.1098/rspb.2008.1082
\bibitem{HilSig10}
Hilbe C, Sigmund K (2010) Incentives and opportunism: from the carrot to the stick. Proc R Soc B 277:2427--2433.
doi:10.1098/rspb.2010.0065
\bibitem{Sutter10}
Sutter M, Haigner S, Kocher MG (2010) Choosing the carrot or the stick? Endogenous institutional choice in social dilemma situations. Rev Econ Stud 77:1540--1566. 
doi:10.1111/j.1467-937X.2010.00608.x
\bibitem{SzoEtal11}
Szolnoki A, Szab{\rm \'{o}} G, Czak{\rm \'{o}} L (2011) Competition of individual and institutional punishments in spatial public goods games. Phys Rev E 84:046106.
doi:10.1103/PhysRevE.84.046106
\bibitem{SasEtal12}
Sasaki T, Br\"{a}nnstr\"{o}m \AA, Dieckmann U, Sigmund K (2012) The take-it-or-leave-it option allows small penalties to overcome social dilemmas. Proc Natl Acad Sci USA 109:1165--1169.
doi:10.1073/pnas.1115219109
\bibitem{PanBo04}
Panchanathan K, Boyd R (2004) Indirect reciprocity can stabilize cooperation without the second-order free rider problem. Nature 432: 499--502. 
doi:10.1038/nature02978
\bibitem{Sky04}
Skyrms B (2004) The Stag Hunt and the Evolution of Social Structure. Cambridge University Press, Cambridge, UK
\bibitem{Akt04}
Aktipis CA (2004) Know when to walk away: contingent movement and the evolution of cooperation. J Theor Biol 231:249--260.
doi:10.1016/j.jtbi.2004.06.020
\bibitem{OrbDaw93}
Orbell JM, Dawes RM (1993) Social welfare, cooperators' advantage, and the option of not playing the game.  Am Soc Rev 58:787--800.
doi:10.2307/2095951
\bibitem{BatKit95}
Batali J, Kitcher P (1995) Evolution of altruism in optional and compulsory games. J Theor Biol 175:161--171.
doi:10.1006/jtbi.1995.0128
\bibitem{HauEtal02a}
Hauert C, De Monte S, Hofbauer J, Sigmund K (2002) Volunteering as Red Queen mechanism for cooperation in public goods games. Science 296:1129--1132.
doi:10.1126/science.1070582
\bibitem{HauEtal02b}
Hauert C, De Monte S, Hofbauer J, Sigmund K (2002) Replicator dynamics for optional public good games. J Theor Biol 218:187--194.
doi:10.1006/jtbi.2002.3067
\bibitem{SemEtal03}
Semmann D, Krambeck HJ, Milinski M (2003) Volunteering leads to rock-paper-scissors dynamics in a public goods game. Nature 425:390--393.
doi:10.1038/nature01986
\bibitem{MatBoy09}
Mathew S, Boyd R (2009) When does optional participation allow the evolution of cooperation. Proc R Soc Lond B 276:1167--1174.
doi:10.1098/rspb.2008.1623
\bibitem{IzqEtal10}
Izquierdo SS, Izquierdo LR, Vega-Redondo F (2010) The option to leave: conditional dissociation in the evolution of cooperation. J Theor Biol 267:76--84.
doi:10.1016/j.jtbi.2010.07.039
\bibitem{CasTor10}
Castro L, Toro MA (2010) Iterated prisoner's dilemma in an asocial world dominated by loners, not by defectors. Theor Popul Biol 74:1--5.
doi:10.1016/j.tpb.2008.04.001
\bibitem{SasEtal07}
Sasaki T, Okada I, Unemi T (2007) Probabilistic participation in public goods games. Proc R Soc B 274:2639--2642.
doi:10.1098/rspb.2007.0673
\bibitem{XuEtal10}
Xu ZJ, Wang Z, Zhang LZ (2010) Bounded rationality in volunteering public goods games. J Theor Biol 264:19--23.
doi:10.1016/j.jtbi.2010.01.025
\bibitem{GarTra12}
García J, Traulsen A (2012) Leaving the loners alone: evolution of cooperation in the presence of antisocial punishment. J Theor Biol 307:168--173. 
doi:10.1016/j.jtbi.2012.05.011
\bibitem{ZhongEtal13}
Zhong LX, Xu WJ, Shi YD, Qiu T (2013) Coupled dynamics of mobility and pattern formation in optional public goods games. Chaos Solitons Fractals 47:18--26. 
doi:10.1016/j.chaos.2012.11.012
\bibitem{Fow05}
Fowler J (2005) Altruistic punishment and the origin of cooperation. Proc Natl Acad Sci USA 102:7047--7049.
doi:10.1073/pnas.0500938102
\bibitem{BraEtal06}
Brandt H, Hauert C, Sigmund K (2006) Punishing and abstaining for public goods. Proc Natl Acad Sci USA 103:495--497.
doi:10.1073/pnas.0507229103
\bibitem{HauEtal07}
Hauert C, Traulsen A, Brandt H, Nowak MA, Sigmund K (2007) Via freedom to coercion: the emergence of costly punishment. Science 316:1905--1907.
doi:10.1126/science.1141588
\bibitem{DesEtal09}
De Silva H, Hauert C, Traulsen A, Sigmund K (2009) Freedom, enforcement, and the social dilemma of strong altruism. J Evol Econ 20:203--217.
doi:10.1007/s00191-009-0162-8
\bibitem{SigEtal10}
Sigmund K, De Silva H, Traulsen A, Hauert C (2010) Social learning promotes institutions for governing the commons. Nature 466:861--863.
doi:10.1038/nature09203
\bibitem{SigEtal11}
Sigmund K, Hauert C, Traulsen A, De Silva H (2011) Social control and the social contract: the emergence of sanctioning systems for collective action. Dyn Games Appl 1:149--171.
doi:10.1007/s13235-010-0001-4
\bibitem{TRM12}
Traulsen A, R\"{o}hl T, Milinski M (2012) An economic experiment reveals that humans prefer pool punishment to maintain the commons. Proc R Soc B 279:3716--3721. 
doi:10.1098/rspb.2012.0937
\bibitem{Yam86}
Yamagishi T (1986) The provision of a sanctioning system as a public good. J Pers Soc Psychol 51:110--116. 
doi:10.1037/0022-3514.51.1.110
\bibitem{Perc12}
Perc M (2012) Sustainable institutionalized punishment requires elimination of second-order free-riders. Sci Rep 2:344. 
doi:10.1038/srep00344
\bibitem{CreEtal11}
Cressman R, Song JW, Zhang BY, Tao Y (2011) Cooperation and evolutionary dynamics in the public goods game with institutional incentives. J Theor Biol. 299:144--151. doi:10.1016/j.jtbi.2011.07.030
\bibitem{BalGro11}
Baldassarri D, Grossman G (2011) Centralized sanctioning and legitimate authority promote cooperation in humans. Proc Nat Acad Sci USA 108:11023--11026.
doi:10.1073/pnas.1105456108
\bibitem{IsaRan12}
Isakov A, Rand DG (2012) The evolution of coercive institutional punishment. Dyn Games Appl 2:97--109. 
doi:10.1007/s13235-011-0020-9
\bibitem{AndGee12}
Andreoni J, Gee LK (2012) Gun for hire: delegated enforcement and peer punishment in public goods provision. J Public Econ 96:1036--1046. 
doi:10.1016/j.jpubeco.2012.08.003
\bibitem{SasUne11}
Sasaki T, Unemi T (2011) Replicator dynamics in public goods games with reward funds. J Theor Biol 287:109--114.
doi:10.1016/j.jtbi.2011.07.026
\bibitem{HofSig98}
Hofbauer J, Sigmund K (1998) Evolutionary Games and Population Dynamics. Cambridge University Press, Cambridge, UK
\bibitem{Sugden86}
Sugden, R (1986) The Economics of Rights, Co-operation and Welfare. Blackwell, Oxford, UK
\bibitem{Kim96}
Kim Y (1996) Equilibrium selection in $n$-person coordination games. Games Econ Behav 15:203--227. 
doi:10.1006/game.1996.0066
\bibitem{Hof99}
Hofbauer J (1999) The spatially dominant equilibrium of a game. Ann Oper Res 89:233--251. 
doi:10.1023/A:1018979708014
\bibitem{GoVe05}
Goyal S, Vega-Redondo F (2005) Network formation and social coordination. Games Econ Behav 50:178--207. 
doi:10.1016/j.geb.2004.01.005
\bibitem{KMR93}
Kandori M, Mailath G, Rob R (1993) Learning, mutation, and long-run equilibria in games. Econometrica 61:29--56. 
doi:10.2307/2951777
\bibitem{Young93}
Young PH (1993) The evolution of conventions. Econometrica 61:57--84. 
doi:10.2307/2951778
\bibitem{Elli00}
Ellison G (2000) Basins of attraction, long-run stochastic stability, and the speed of step-by-step evolution. Rev Econ Stud 67:17--45. 
doi:10.1111/1467-937X.00119
\bibitem{HarSel88}
Harsanyi JC, Selten R (1988) A General Theory of Equilibrium Selection in Games. MIT Press, Cambridge, MA
\bibitem{Oech97}
Oechssler J (1997) Decentralization and the coordination problem. J Econ Behav Organ 32:119--135. 
doi:10.1016/S0167-2681(96)00022-4
\bibitem{Elli00}
Colman AM (2006) The puzzle of cooperation. Nature 440:744--745. 
doi:10.1038/440744b
\bibitem{HerEtal08}
Herrmann B, Th\"{o}ni C, G\"{a}chter S (2008) Antisocial punishment across societies. Science 319:1362--1367.
doi:10.1126/science.1153808
\bibitem{RanEtal10}
Rand DG, Armao JJ, Nakamaru M, Ohtsuki H (2010) Anti-social punishment can prevent the co-evolution of punishment and cooperation. J Theor Biol 265:624--632.
doi:10.1016/j.jtbi.2010.06.010
\bibitem{RanNow11}
Rand DG, Nowak MA (2011) The evolution of antisocial punishment in optional public goods games. Nature Communications 2:434.
doi:10.1038/ncomms1442

\end{thebibliography}


\end{document}